\documentclass{iopart}
\usepackage{epsfig}

\begin{document} \title{A Simulation of the LISA Data Stream from Galactic White Dwarf Binaries}

\author{M J Benacquista, J DeGoes, D Lunder}

\address{Dept.\ of Sciences, Montana State University-Billings, Billings, MT 59101}
\ead{benacquista@msubillings.edu}

\begin{abstract} Gravitational radiation from the galactic population of white dwarf binaries is expected to produce a background signal in the LISA frequency band. At frequencies below 1 mHz, this signal is expected to be confusion-limited and has been approximated as gaussian noise. At frequencies above about 5 mHz, the signal will consist of separable individual sources. We have produced a simulation of the LISA data stream from a population of 90k galactic binaries in the frequency range between 1 - 5 mHz. This signal is compared with the simulated signal from globular cluster populations of binaries. Notable features of the simulation as well as potential data analysis schemes for extracting information are presented. \end{abstract}


\section{Introduction}

The Laser Interferometer Space Antenna (LISA) is a planned space-based gravitational wave detector. It will be sensitive to gravitational radiation in the frequency band from $10^{-5} \-- 1~{\rm Hz}$, with sources ranging from supermassive black hole binaries to stellar mass compact binaries. One of the most abundant sources will be the galactic population of close white dwarf binaries. These sources will dominate the LISA signal from 0.1 mHz up to about 3 mHz and are expected to produce a confusion limited background below $\sim 1~{\rm mHz}$. The signal will transition to individually resolved sources above $\sim 5~{\rm mHz}$. Data analysis strategies to accommodate the close white dwarf binaries vary according to the nature of the signal. Below 1 mHz, the signal will be isotropic and effectively gaussian with $> 100$ sources per frequency bin for a one year observation. Above 5 mHz, the signal will individual sources widely separated in frequency. The transition region between 1 and 5 mHz presents unique challenges for data analysis. In addition, sources of gravitational radiation in globular clusters have the greatest angular resolution in this transition zone. It is therefore of use to develop a simulation of the LISA data stream from galactic white dwarf binaries in the transition zone between 1 and 5 mHz test data analysis techniques for handling the foreground signal while identifying other signals of interest such as globular cluster binaries or high mass ratio inspirals into supermassive black holes. We have generated just such a simulation with 90,000 white dwarf binaies and describe in Section~\ref{popsyn} the population synthesis and in Section~\ref{datstream} the data stream simulation. The simulated data stream from select globular clusters is described in Section~\ref{globs}.

\section{Population Synthesis}\label{popsyn}

The population of galactic white dwarf binaries is synthesized using a phenomenological approach. We assume a uniform birthrate over the past $10^{10}$ yr, and then assign a binary type at birth according the simulations of Nelemans \etal (2001). The birth of the binary is considered to occur after the second white dwarf is formed. The binary types are combinations of four basic white dwarf types \-- Helium, hybrid Carbon Oxygen (with a thick Helium envelope), Carbon Oxygen, and Oxygen Neon. The binary types and their probabilities at birth are shown in Table~\ref{bintypes}
\begin{table}
\label{bintypes}
\caption{Probabilities of binary types at birth.}
\begin{indented}
\item[]\begin{tabular}{@{}lll}
\br
Probability & Primary & Secondary \\
\mr
0.54 & He & He \\
0.2225 & CO & CO \\
0.112 & CO & He \\
0.048 & He & CO \\
0.028 & HyCO & He \\
0.015 & HyCO & CO \\
0.0125 & HyCO & HyCO \\
0.012 & He & HyCO \\
0.005 & ONe & He \\
0.005 & ONe & CO \\
\br
\end{tabular}
\end{indented}
\end{table}
Once the binary types have been determined, individual masses are assigned to each white dwarf based upon its type. These are drawn from mass distributions derived from Iben and Tutukov (1986). The orbital period, $Q$ (in hours) at birth is chosen by assuming a normal distribution with unit variance in $\log{(Q)}$ with the mean period $\bar{Q}$ related to the mass of the secondary using $m = 0.0625 \log{(\bar{Q})} + 0.2625$ if the secondary is a Helium white dwarf and $\log{(\bar{Q})} = 0.4$ if it is any other type. Finally, the binary is assigned a birth date between the present and $10^{10}$ years ago. The binary is allowed to evolve under the influence of gravitational radiation to determine the present orbital period ($P$) using
\begin{equation}
P = \left(Q^{8/3} - \frac{8}{3} k_0 t\right)^{3/8}
\end{equation}
with
\begin{equation}
k_0 = \frac{96}{5}\left(2\pi\right)^{8/3}\frac{G^{5/3}M_1M_2}{c^5\left(M_1+M_2\right)^{1/3}}
\end{equation}
where $M_1$ and $M_2$ are the masses of the white dwarfs (Peters and Mathews, 1963). Finally, the binary is retained as part of the synthesized population if its orbital period is less than $2000$ s and greater than the period at Roche lobe contact.

Once the masses and orbital period of the binary have been established, it is then assigned a position in the galaxy according to the probability distribution functions in the cylindrical coordinates $R$ and $z$ given by:
\begin{eqnarray}
P(R) =\frac{1}{R_0} \e^{-R/R_0} \\
P(z) = \frac{1}{2z_0} {\rm sech}^2{(z/z_0)},
\end{eqnarray}
where $R_0 = 2500~{\rm pc}$ and $z_0 = 200~{\rm pc}$. The angular coordinate is chosen from a flat distribution and the coordinates are converted to heliocentric spherical polar coordinates, $(r,\theta_s,\phi_s)$, with $\phi_s$ measured from the vernal equinox and $\theta_s$ measured from the normal to the ecliptic. The orientation of the binary is described by the direction of the angular momentum vector. The angular coordinates of this vector are assigned using flat distribution functions and are $(\theta_L,\phi_L)$ in the same heliocentric coordinate system. Finally, the orbital phase of the binary ($\phi_0$) is also chosen randomly. The population of binaries is then described by 9 parameters \--- the masses of the components ($M_1$ and $M_2$), the orbital period ($P$), the position in heliocentric coordinates ($r, \theta_s, \phi_s$), the orientation ($\theta_L, \phi_L$), and the orbital phase ($\phi_0$).

\begin{figure}
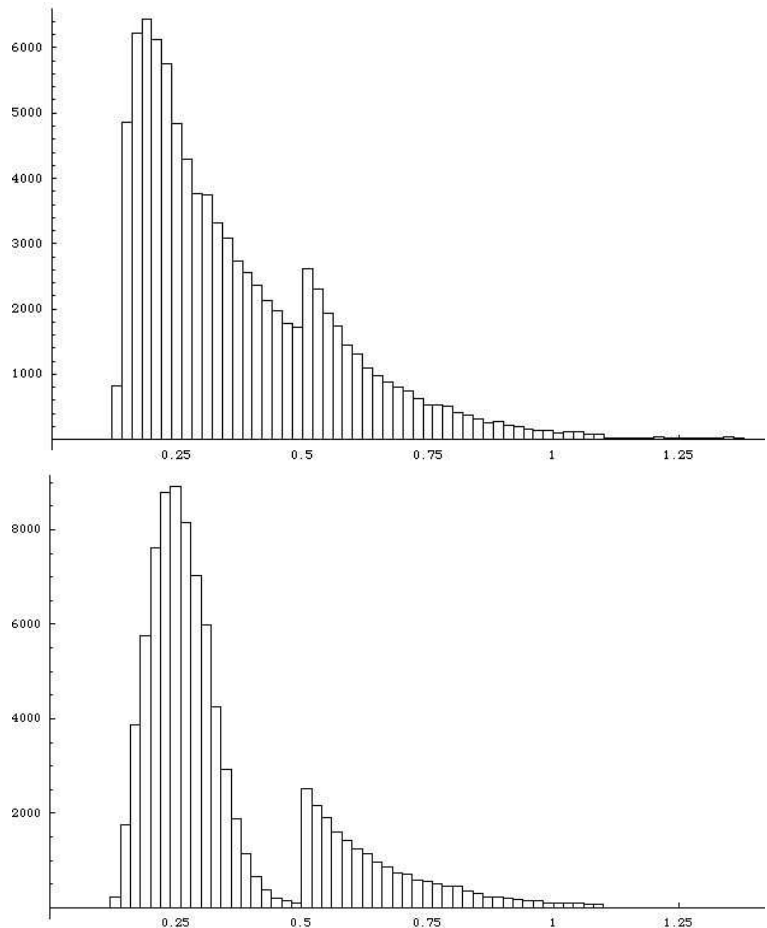

\begin{center}
\epsfig{file=Mass_1_hist.epsf,width=10cm}
\epsfig{file=Mass_2_hist.epsf,width=10cm}
\end{center}
\caption{Distribution of the masses of the white dwarfs in solar masses. The upper histogram shows the masses of the first white dwarf formed while the lower one gives the second white dwarf formed.}
\label{massdist}
\end{figure}

\begin{figure}
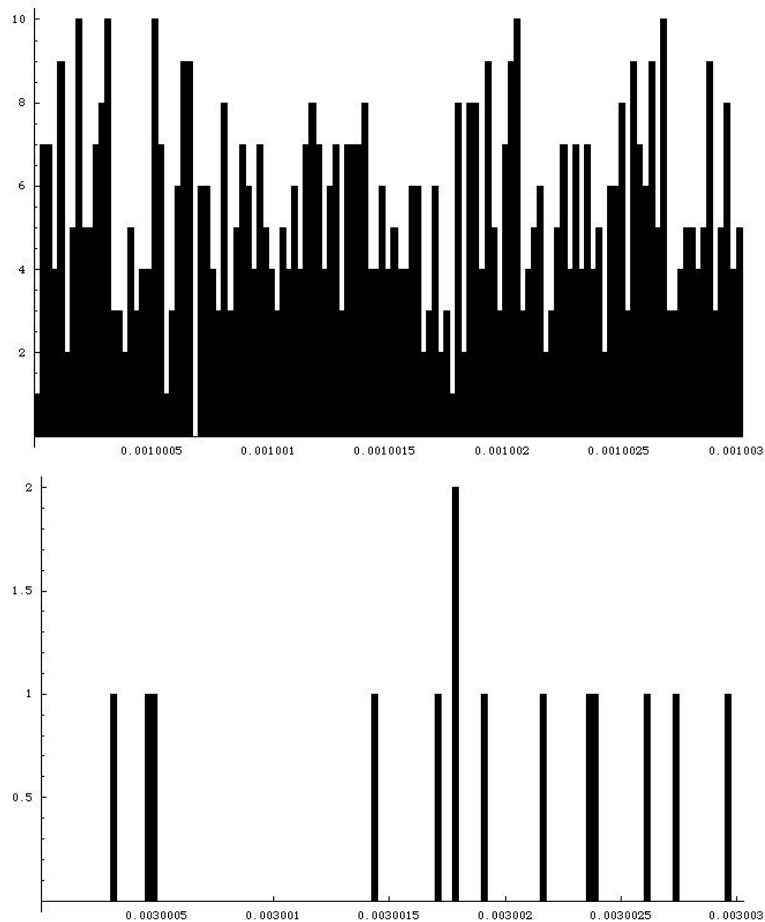

\begin{center}
\epsfig{file=First100bins.epsf,width=10cm}
\epsfig{file=Mid100bins.epsf,width=10cm}
\end{center}
\caption{Number of binaries per $1~{\rm yr}^{-1}$ frequency bin. The upper histogram shows the first 100 bins above 1 mHz while the lower one shows the first 100 bins above 3 mHz.}
\label{freqdist}
\end{figure}

The number of binaries to include in the population is chosen to be 90,000 which is close to the number determined by using the population estimates of Hils, Bender, and Webbink (1990) and (following their lead) reducing this number by a factor of 10. The distribution of the masses of the binaries are shown in Figure~\ref{massdist}, and the number of sources per frequency bin for selected frequencies are shown in Figure~\ref{freqdist}.

\section{Data Stream Simulation}\label{datstream}

\begin{figure}
\begin{center}
\epsfig{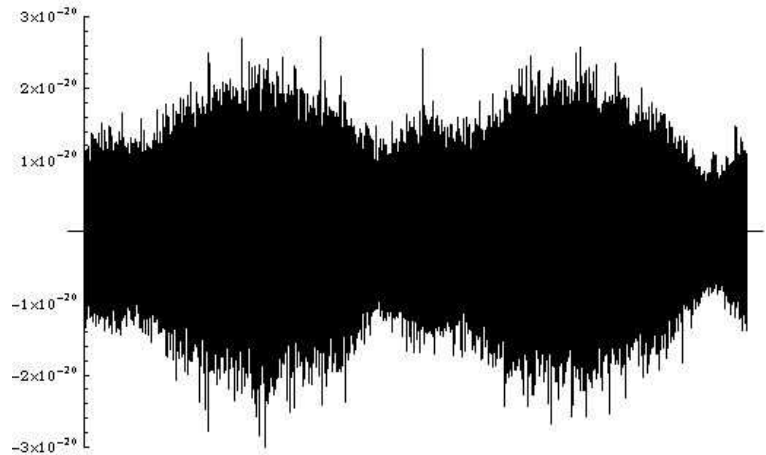}
\end{center}
\caption{Full year data stream of $h_1$ strain amplitude. The time axis covers $32 \times 10^6$ s and the strain amplitude is less than $3 \times 10^{-20}$.}
\label{datastream}
\end{figure}

The LISA data stream simulated from the population of binaries is calculated using the long wavelength approximation of Cutler (1998). Cornish and Rubbo (2003) have shown that this low frequency approximation will suffice for frequencies below $\sim 5$ mHz. We have calculated the data stream for both the $h_I$ and $h_{II}$ configurations of Cutler and including both the Doppler phase ($\varphi_D$) and the polarization phase ($\varphi_p$). We have chosen for the initial configuration of LISA to be $\bar{\phi}_0 = 0$ and $\alpha_0 = 0$ which correspond to LISA starting at the vernal equinox at $t = 0$. The binaries are assumed to have been circularized and therefore emitting monochromatic gravitational radiation with frequency $f = 2/P$. The data stream is calculated for each binary as a double-precision strain amplitude measured every second for $32 \times 10^6$ seconds. These data streams have then been combined to form a single stream for all 90,000 binaries. The data stream for $h_I$ is shown in Figure~\ref{datastream}. It is of interest to note that the amplitude of the signal from the galactic binaries can be seen to vary annually as the antenna pattern of LISA sweeps through the plane of the galaxy, thus showing that the anisotropy of the binary distribution is measurable in this frequency band.

\section{Globular Cluster Simulations}\label{globs}

\begin{figure}
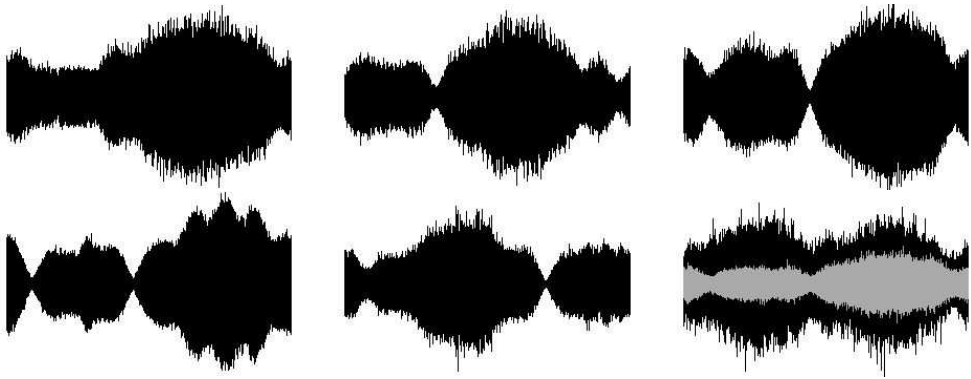

\begin{center}
\centerline{\epsfig{file=ngc104.epsf,width=4cm}\hspace{0.5cm}\epsfig{file=ngc5139.epsf,width=4cm}\hspace{0.5cm}\epsfig{file=ngc6397.epsf,width=4cm}}
\centerline{\epsfig{file=ngc6752.epsf,width=4cm}\hspace{0.5cm}\epsfig{file=ngc7078.epsf,width=4cm}\hspace{0.5cm}\epsfig{file=gcanddisk.epsf,width=4cm}}\end{center}
\caption{Data streams for NGC 104, NGC 5139, NGC 6397, NGC 6752, NGC 7078, and the five globular clusters (in grey) compared with the disk signal (in black). The data streams from the individual globular clusters are plotted at different scales to show the details. The actual amplitudes range from $2 \times 10^{-22}$ to $1.1 \times 10^{-21}$.}
\label{globdat}
\end{figure}

Although there have been a few simulations of the white dwarf binary population of globular cluster (Hurley and Shara 2003), a detailed synthesis of such a population is very difficult due to the additional complication of the internal dynamics of a globular cluster. Consequently, we have simply chosen to use binaries from our galactic population synthesis to simulate the signal from globular cluster binaries. Since the globular cluster binaries are more localized in space, the time variation of their signal should be different than that from galactic binaries and this is borne out in the simulations performed for five individual globular clusters containing 10 binaries each. The globular cluster populations were simulated by choosing 10 binaries from the population synthesis described in Section~\ref{popsyn} with their positions chosen from a three-dimensional normal distribution centered on the position of the globular cluster with a variance given by the half-mass radius. The globular cluster data was taken from Harris (1996). The datastreams from these globular clusters and a combined data stream are shown in Figure~\ref{globdat}. Because the globular clusters are much more localized sources, their signals can be reduced to zero as it passes through a null in the antenna pattern of LISA. Searching for such nulls may provide a technique for identifying globular cluster binaries.

\section{Conclusion}

We have synthesized a population of close white dwarf binaries in the galactic disk and used this population to generate a simulated data stream from LISA. This simulation does not include the instrument noise for LISA, but noise can be included in the signal by simply adding the appropriately chosen noise spectrum to the signal. In this way, our simulated data stream can be used to compare the effect of different instrumental noise spectra on the ability of different data analysis techniques to handle the signal from the galactic population of white dwarf binaries. The simulation can also be combined with simulated signals for other sources of interest (as we have done with globular cluster sources), to test different data analysis schemes for extracting these other signals.

We anticipate improving the simulation by including white dwarf binaries with orbital periods greater than $2000$ s and incorporating the general response function of Cornish and Rubbo (2003) for binaries with frequencies higher than $\sim 3$ mHz. In the meantime, we will make the simulations freely available on CD for anyone interested in using it for testing data analysis schemes.

\ack

The authors acknowledge support from NASA Cooperative Agreement NCC5-579 and the Montana Space Grant Consortium.

\References
\item[] Cornish N J and Rubbo L J 2003 {\it Phys. Rev} D {\bf 67} 022001
\item[] Cutler C 1998 {\it Phys. Rev} D {\bf 57} 7089
\item[] Harris W E 1996 {\it Astron. J.} {\bf 112} 1487
\item[] Hils D, Bender P L and Webbink R F 1990 {\it Astrophys. J.} {\bf 360} 75
\item[] Hurley J R and Shara M M 2003 {\it Astrophys. J.} {\bf 589} 179
\item[] Iben I Jr. and Tutukov A V 1986 {\it Astrophys. J.} {\bf 311} 753
\item[] Nelemans G, Yungelson L R, Portegies Zwart S F and Verbuunt F 2001 {\it Astron. Astrophys.} {\bf 365} 491
\item[] Peters P C and Mathews J 1963 {\it Phys. Rev.} {\bf 131} 435
\endrefs

\end{document}